\documentclass[twocolumn]{aastex63}
\begin{document}
\shorttitle{Investigation of Primordial Cluster Pair}

\shortauthors{Hu, Q.S. et al.}

\title{\textbf{An In-depth} Investigation of the Primordial Cluster Pair ASCC~19 and ASCC~21}

\author{Qingshun Hu}
\email{qingshun0801@163.com}
\affil{School of Physics and Astronomy, China West Normal University, No. 1 Shida Road, Nanchong 637002, People's Republic of China}

\author{Yuting Li}
\affil{School of Physics and Astronomy, China West Normal University, No. 1 Shida Road, Nanchong 637002, People's Republic of China}

\author{Mingfeng Qin}
\affil{Institute for Frontiers in Astronomy and Astrophysics, Beijing Normal University, Beijing 102206, People's Republic of China}
\affil{School of Physics and Astronomy, Beijing Normal University, Beijing 100875, People's Republic of China}

\author{Chenglong Lv}
\affil{Xinjiang Astronomical Observatory, Chinese Academy of Sciences, No. 150, Science 1 Street, Urumqi, Xinjiang 830011, People's Republic of China}

\author{Yang Pan}
\affil{School of Physics and Astronomy, China West Normal University, No. 1 Shida Road, Nanchong 637002, People's Republic of China}

\author{Yangping Luo}
\affil{School of Physics and Astronomy, China West Normal University, No. 1 Shida Road, Nanchong 637002, People's Republic of China}

\author{Shuo Ma}
\affil{Department of Astronomy, Xiamen University, Xiamen, Fujian 361005, People's Republic of China}

\begin{abstract}
	
	Utilizing \texttt{Gaia} data from the literature, we report a new young ($\sim$8.9~Myr) cluster pair, ASCC~19 and ASCC~21, located near the Orion star-forming complex. The clusters are separated by a 3D distance of ~27.00~$\pm$~7.51~pc. Both clusters share a common age (Log(age)~=~6.95~$\pm$~0.05), similar radial velocities ($R_{v}$~=~21.34~$\pm$~4.47~km·s$^{-1}$ for ASCC~19 and $R_{v}$~=~20.05~$\pm$~3.86~km·s$^{-1}$ for ASCC~21), and comparable metallicities ([Fe/H]~=~$-$0.14~$\pm$~0.25~dex for ASCC~19 and [Fe/H]~=~$-$0.12~$\pm$~0.04~dex for ASCC~21, from LAMOST-DR11). These similarities suggest that the clusters likely originated from the fragmentation of the same molecular cloud, forming a primordial cluster pair. Furthermore, the formation of the two clusters is attributed to the coalescence of multiple subclusters, as inferred from the distribution analysis between metal abundances and distances to clusters' centers. Neither cluster shows significant mass segregation. Their members with radial velocities exceeding 100~km·s$^{-1}$ are young variables. Additionally, a tidal interaction between the clusters is observed. Comparisons of the Roche radius with tidal radii, and velocity difference with orbital velocity, suggest that the pair is an unbound system, that is, a double cluster. Finally, orbital motion simulations show that the clusters will not merge into a single system.

\end{abstract}



\section{Introduction}

Open clusters (OCs) are usually found not only in isolation within the galactic disk but also in pairs within the disk. \citet{dela09} found that the fraction of OC pairs at the Solar Circle is comparable to that in the Magellanic Cloud, approximately 10\% \citep{bhat88,hatz90} or 12\% \citep{piet00, dieb02}. Cluster pairs generally consist of binary and double clusters \citep{dela09}. Binary clusters are gravitationally bound pairs, while double clusters are unbound pairs.

Several formation scenarios have been proposed for cluster pairs. First, binary clusters may form through tidal capture, as suggested by \citet{dela09}. These pairs are usually not primordial, exhibiting common motion velocities but differing chemical compositions, and ages. Second, resonances induced by the non-axisymmetric component of the Galactic potential, such as the Galactic bar or spiral arms \citep[e.g.][]{dehn98, desi04, fama05, quil05, chak07}, can facilitate the formation of double clusters. Third, cluster pairs may form simultaneously within the same giant molecular cloud, resulting in similar ages, kinematics, and chemical compositions, as described by \citet{dela09}. These cluster pairs can be called primordial cluster pairs. Fourth, cluster pairs can form sequentially, with supernova shocks or massive stellar winds from one cluster triggering the formation of another cluster from adjacent clouds, a mechanism similar to the formation of globular clusters \citep{brow95, good97}. Fifth, optical double clusters usually form through chance alignments or hyperbolic encounters.

Nevertheless, double or binary OCs are generally not stable over long periods of time. The components of a double cluster generally undergo a hyperbolic flyby, after which they separate. For binary clusters, numerical simulations \citep{port07, dela10, priy16} demonstrate that they, often observed only in their early stages, survive for short times. Binary clusters follow two evolutionary paths: either they become two completely independent, separated, non-interacting clusters due to tidal disruption and ionization, or they merge into a single object \citep{dela10}. In addition, binary clusters may also be disrupted during mutual tidal interactions (shredded secondaries) \citep{dela10}. Therefore, detecting double or binary OCs is challenging.

Before \texttt{Gaia}, limited data made double or binary cluster identification difficult. Despite the challenge, some studies have identified cluster pairs. For instance, \citet{pavl89} and \citet{subr95} identified 5 possible cluster groups and 18 possible cluster pairs, respectively. Later, \citet{dela09} identified 34 OC pairs using data from NCOVOCC \citep{dias02} and WEBDA \citep{merm03}. \citet{conr17} singled out 14 binary clusters from the catalogues of \citet{khar04} and \citet{khar05}. Subsequently, additional clusters were detected with more comprehensive post-\texttt{Gaia} surveys. Recently, \citet{song22} and \citet{li24} respectively identified 14 binary OC candidates based on the catalog of \citet{cant20} and 13 newly close binary OCs via the catalog of \citet{hunt23}, both derived from Gaia data. Similarly, \citet{casa21a} identified 22 new binary or multiple OCs from various OC catalogs \citep{khar13, cant18, cant20, bica19, liu19, sim19, cast20}. \citet{casa21b} also discovered a new binary OC, named Casado~9 and Casado~10, using Gaia~DR2 data.

\begin{figure*}[ht]
	\centering
	\includegraphics[width=120mm]{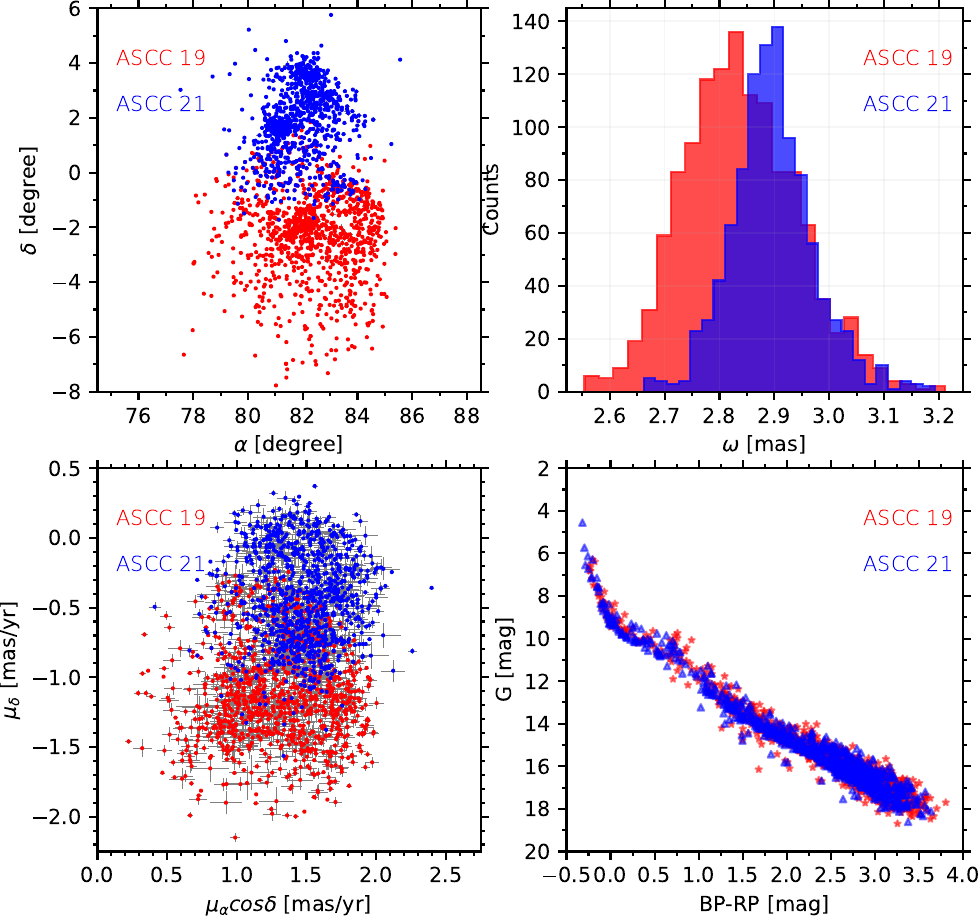}
	\caption{Member distributions of ASCC~19 and ASCC~21 in multi-dimensional parameter spaces. The figure includes only members with a membership probability of 1. The members of ASCC~19 are marked by red dots or filled pentagrams, with those of ASCC~21 by blue dots or filled triangles. \textit{Top left panel}: Distribution of the members of both clusters in a two-dimensional (2D) celestial coordinate system. Gray bars represent the errors of $\alpha$ and $\delta$. \textit{Top right panel}: Parallax histograms of the member stars for both clusters. \textit{Bottom left panel}: Member distribution of both clusters in proper motion space. Gray bars represent the errors of proper motion in both directions. \textit{Bottom right panel}: Color Magnitude Diagram (CMD) for both clusters. All member data are sourced from \citet{van23}.}
	\label{Double_clusters}
\end{figure*}

Although numerous studies have reported plenty of cluster pairs, it is still possible that many more, especially primordial cluster pairs, remain to be identified within the Milky Way. This is because the simultaneous fragmenting and collapsing of giant molecular clouds can form multiple clusters, possibly leading to the formation of a certain number of primordial cluster pairs. The availability of precise \texttt{Gaia} data has provided new opportunities for discovering primordial cluster pairs, particularly from recently published catalogs, such as that by \citet{van23}, based on the third Gaia Data Release (Gaia~DR3\footnote{\url{https://gea.esac.esa.int/archive/}} \citep{gaia23}). Additionally, \citet{hunt24} precisely classified clusters and provided a list of true OCs. This can contribute to the accurate identification of cluster pairs. In this study, we searched for nearby OCs and discovered a new primordial cluster pair, ASCC~19 and ASCC~21, which had not been previously reported.

The rest of this paper is organized as follows: Sect.~2 describes our data selection process; Sect.~3 analyzes the cluster pair consisting of ASCC~19 and ASCC~21 and determines that it is a primordial cluster pair; Sect.~4 investigates the properties of the pair; Sect.~5 discusses whether the primordial cluster pair is bound; Sect.~6 presents a summary.

\begin{figure}[ht]
	\centering
	\includegraphics[width=80mm]{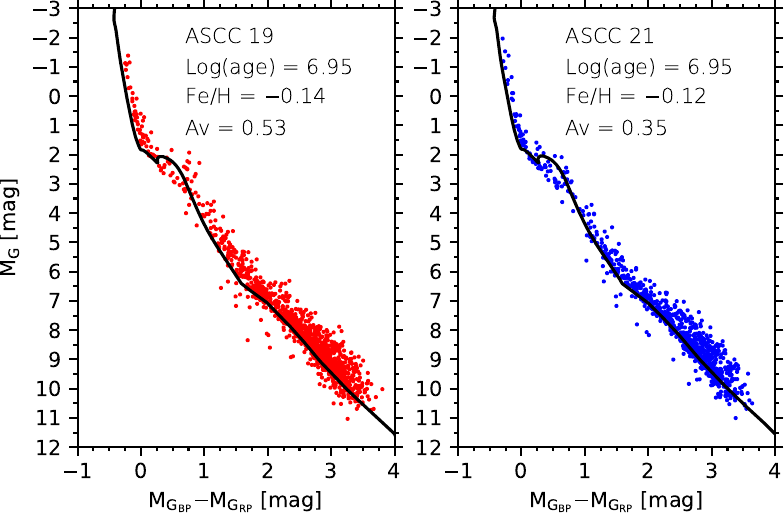}
	\caption{\textit{Left panel}: CMD of ASCC~19. \textit{Right panel}: CMD of ASCC~21. The solid black lines represent the isochrones fitted using the \texttt{PARSEC} model. The red and blue dots indicate the members of ASCC~19 and ASCC~21, respectively. All member data are sourced from \citet{van23}.}
	\label{Double_clusters_CMD}
\end{figure}

\section{Data Selection}

\label{data}

\begin{figure}[ht]
	\centering
	\includegraphics[width=80mm]{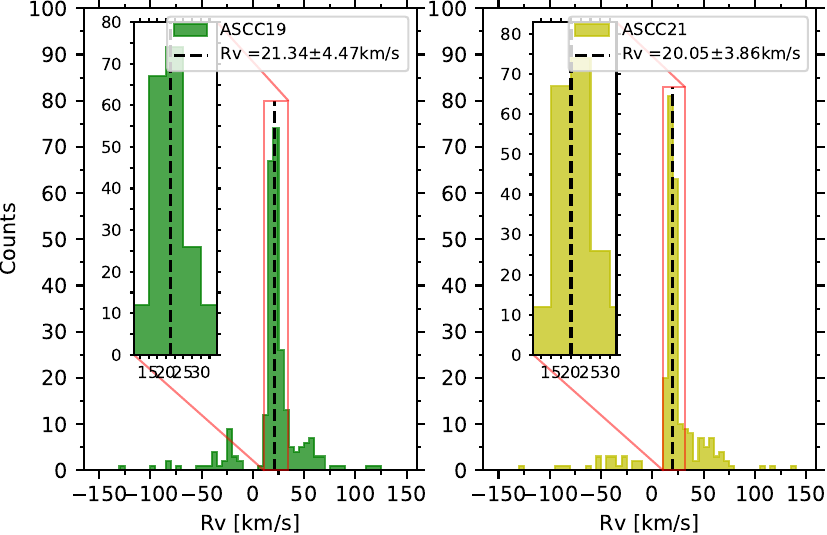}
	\caption{\textit{Left panel}: Radial velocity histogram of ASCC~19. \textit{Right panel}: Radial velocity histogram of ASCC~21. The black dashed lines in each panel denote the median of each distribution. It should be noted that the radial velocity-to-error ratios of the members of both clusters, presented in these histograms, are greater than or equal to 3. The range within three times the median error (Median Absolute Deviation) is marked by the red box in each panel. Radial velocities of the two clusters are from \texttt{Gaia~DR3}.}
	\label{Double_clusters_rv}
\end{figure}

The member stars of ASCC~19 and ASCC~21 were selected from a catalog of 2492 OCs published by \citet{van23} and based on Gaia~DR3 data. This catalog was created, using a deep neural network architecture, with reference to parameters from the OCs catalogs of \citet{cant20} and \citet{cast22}. It provides a more comprehensive list of OC member stars, including many faint stars not present in previous catalogs (see their Fig.~4). ASCC~19 and ASCC~21 were found to contain 3861 and 2709 members, respectively, as shown in Table.~\ref{table:data}. Membership probabilities for both clusters range from 0 to 1. The \texttt{G}-band apparent magnitudes of the member stars of these two open clusters are both brighter than 21~mag. Stars fainter than 19~mag in the \texttt{G}-band usually have large uncertainties \citep{gaia23}. To ensure the reliability of the member stars, we excluded those with a membership probability lower than 1. As a result, ASCC~19 has 1250 true members, and ASCC~21 has 998. After applying the membership probability cut, the \texttt{G}-band apparent magnitudes of the member stars in both open clusters are found to be brighter than 19~mag. It is observed that the number of stars with a membership probability of 1 for both clusters is approximately one-third of that when no membership cut is applied. However, the range of \texttt{G}-band magnitudes has contracted by only 2~mag. Additionally, we did a completeness assessment of the selected cluster member stars, as described in the Appendix~\ref{appendix}, which shows the completeness of the members of ASCC~19 and ASCC~21 are relatively high.

\begin{figure}[ht]
	\centering
	\includegraphics[width=62mm]{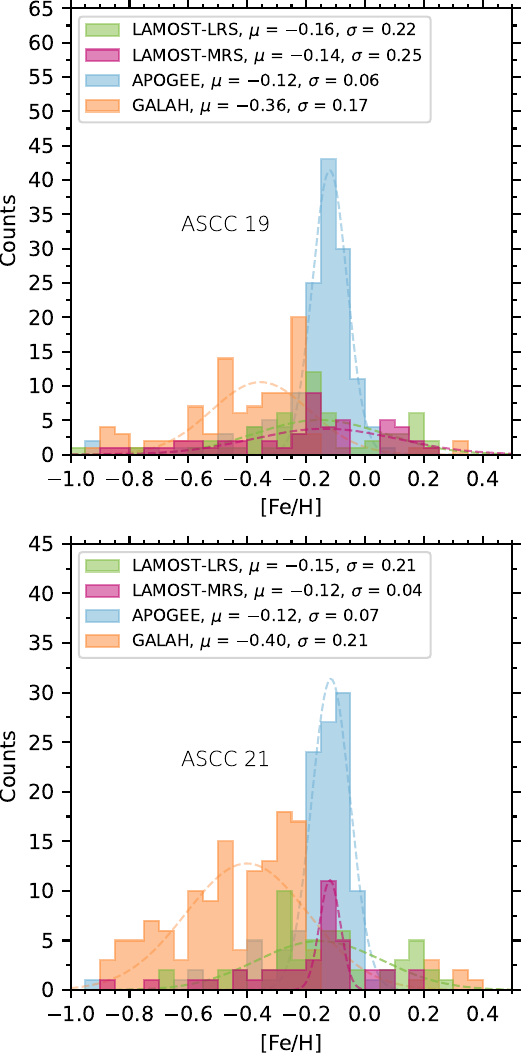}
	\caption{\textit{Top panel}: Histograms of the metallicity ([Fe/H]) of ASCC~19, with the metallicity parameters sourced from the LRS of LAMOST-DR11 (green), the MRS of LAMOST-DR11 (purple), APOGEE-DR17 (blue), and GALAH-DR4 (yellow). \textit{Bottom panel}: Histograms of the metallicity ([Fe/H]) of ASCC~21, with the metallicity parameters derived from the same sources: LRS of LAMOST-DR11, the MRS of LAMOST-DR11, APOGEE-DR17, and GALAH-DR4. The colored dashed lines are Gaussian fitting profiles for the metallicities of the two clusters, with their respective mean and variance being shown in the labels.}
	\label{Double_Fe_H_four}
\end{figure}

\begin{figure}[ht]
	\centering
	\includegraphics[width=88mm]{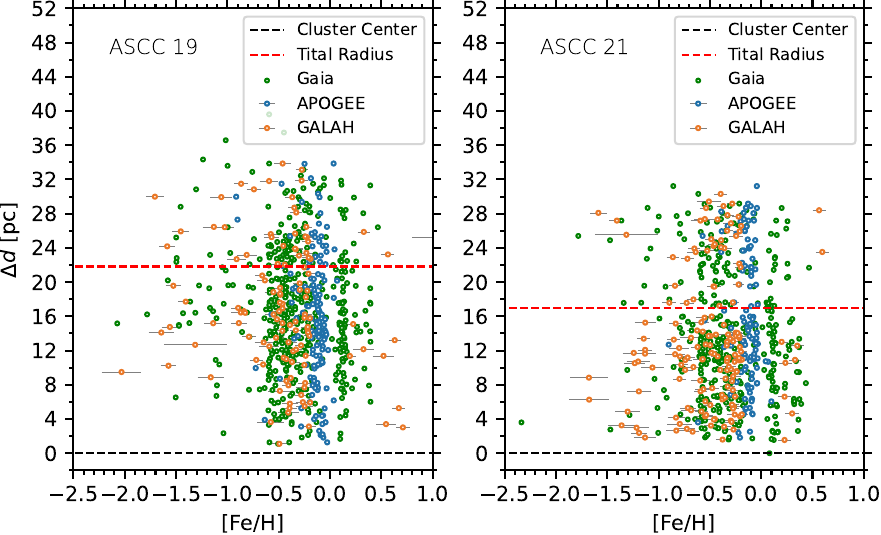}
	\caption{\textit{Left panel}: the distribution of the metallicity [Fe/H] of the members of ASCC~19, derived from Gaia~DR3 (green), APOGEE-DR17 (blue), and GALAH-DR4 (yellow), and the distances ($\Delta d$) to clusters' centers (3D centers). \textit{Right panel}: the distribution of the [Fe/H] of the members of ASCC~21, again from Gaia DR3, APOGEE-DR17, and GALAH-DR4, and the distances ($\Delta d$) to clusters' centers. The black dashed lines represent the 3D centers of the two clusters, with the red dashed lines indicating their tidal radii. The calculation of the tidal radii and the process of deriving distances are detailed in Sec.~\ref{morphology}.}
	\label{Double_Fe_H_d}
\end{figure}

\section{Diagnosis of Cluster Pair}

\begin{table*}[ht]
	\caption{Overall parameters of ASCC~19 and ASCC~21}
	\centering
	
	\label{table:data}
	\begin{tabular}{c c c c}
		\hline\noalign{\smallskip}
		\hline\noalign{\smallskip}
		\hspace{0cm} Parameters & Description & ASCC~19 & ASCC~21 \\
		\hline\noalign{\smallskip}
		$\alpha$ (degree)  & Right ascension  & 82.31~$\pm$~1.07 & 81.91~$\pm$~0.68  \\
		$\delta$ (degree)  & Declination & $-$1.89~$\pm$~1.04 & 1.91~$\pm$~0.96  \\
		$\mu_{\alpha}$ (mas$\cdot$yr$^{-1}$) & Proper motion in right ascension & 1.34~$\pm$~0.25  & 1.48~$\pm$~0.17\\
		$\mu_{\delta}cos\alpha$ (mas$\cdot$yr$^{-1}$)  & Proper motion in declination & $-$1.12~$\pm$~0.24 &  $-$0.44~$\pm$~0.28\\
		$\omega$ (mas)  & Parallax & 2.83~$\pm$~0.07 &  2.89~$\pm$~0.04 \\
		Rv (km$\cdot$s$^{-1}$) & Radial velocity & 21.34~$\pm$~4.47  & 20.05~$\pm$~3.86\\
		$[$Fe$/$H$]$ (dex) & Metallic abundance taken from LAMOST-DR11-LRS & $-$0.16~$\pm$~0.22  & $-$0.15~$\pm$~0.21 \\
		$[$Fe$/$H$]$ (dex) & Metallic abundance taken from LAMOST-DR11-MRS & $-$0.14~$\pm$~0.25  & $-$0.12~$\pm$~0.04 \\
		$[$Fe$/$H$]$ (dex) & Metallic abundance taken from APOGEE-DR17 & $-$0.12~$\pm$~0.06  & $-$0.12~$\pm$~0.07 \\
		$[$Fe$/$H$]$ (dex) & Metallic abundance taken from GALAH-DR4 & $-$0.36~$\pm$~0.17  & $-$0.40~$\pm$~0.21 \\
		Number of all members (-) & Number of members & 3861  & 2709 \\
		Number of real members (-) & Number of members with probability~=~1 & 1250  & 998 \\
		X (pc)  & Spatial position along X axis  & $-$301.70~$\pm$~5.42 & $-$306.73~$\pm$~3.48 \\
		Y (pc)  & Spatial position along Y axis  & $-$140.17~$\pm$~6.90  & $-$118.23~$\pm$~4.11 \\
		Z (pc)  & Spatial position along Z axis  & $-$114.59~$\pm$~5.76  & $-$103.93~$\pm$~5.15 \\
		r$_{h}$ (pc) & Half-mass radius & 15.95~$\pm$~0.01 & 11.62~$\pm$~0.01 \\
		r$_{c}$ (pc) & Core radius & 3.59~$\pm$~0.31 & 2.73~$\pm$~0.29 \\
		r$_{t}$ (pc) & Tidal radius obtained by the intersection point & 21.82 & 16.97 \\
		R$_{t}$ (pc) & Tidal radius estimated by \texttt{von Hoerner formula} & 11.12~$\pm$~0.003 & 10.58~$\pm$~0.002 \\
		Log(age) & Age of clusters & 6.95~$\pm$~0.05 & 6.95~$\pm$~0.05 \\
		R$_{G}$ (pc) & Distance to the Galactic center & 8426.16~$\pm$~5.61 & 8430.04~$\pm$~3.59 \\
		M$_{c}$ (M$_{\odot}$) & Mass of clusters & 796.85 & 686.09 \\
		
		\hline\noalign{\smallskip}
	\end{tabular}
	\flushleft
	
\end{table*}

\begin{figure}[ht]
	\centering
	\includegraphics[width=84mm]{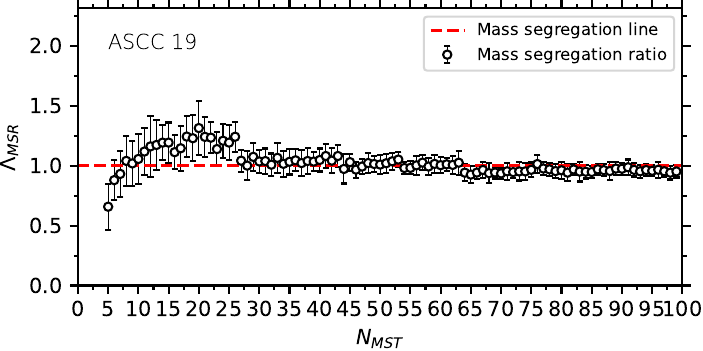}
	\includegraphics[width=84mm]{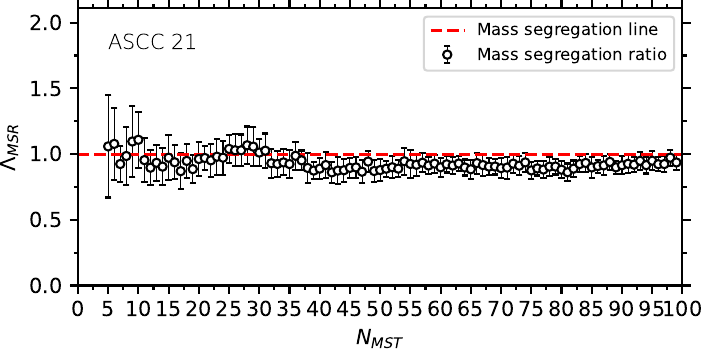}
	\caption{The ``mass segregation ratio" ($\Lambda_{MSR}$) for ASCC~19 and ASCC~21. $N_{MST}$ denotes the ordinal number of the member stars sorted by G-band magnitude, from bright to faint. The parameters are calculated according to the member data sourced from \citet{van23}.}
	\label{mass_segregation}
\end{figure}

This section determined whether ASCC~19 and ASCC~21 are a cluster pair and whether the cluster pair is a primordial cluster pair.

Figure~\ref{Double_clusters} shows the member distributions of ASCC~19 and ASCC~21 across various parameter spaces, including a two-dimensional (2D) celestial coordinate system, parallax, proper motion, and Color Magnitude Diagram (CMD) spaces. We found from the figure that there are overlaps between ASCC~19 and ASCC~21 in the 2D celestial coordinate system, parallax space, and proper motion space, which implies they constitute a cluster pair.

\subsection{Age}\label{Primordial}

\begin{figure*}[ht]
	\centering
	\includegraphics[width=120mm]{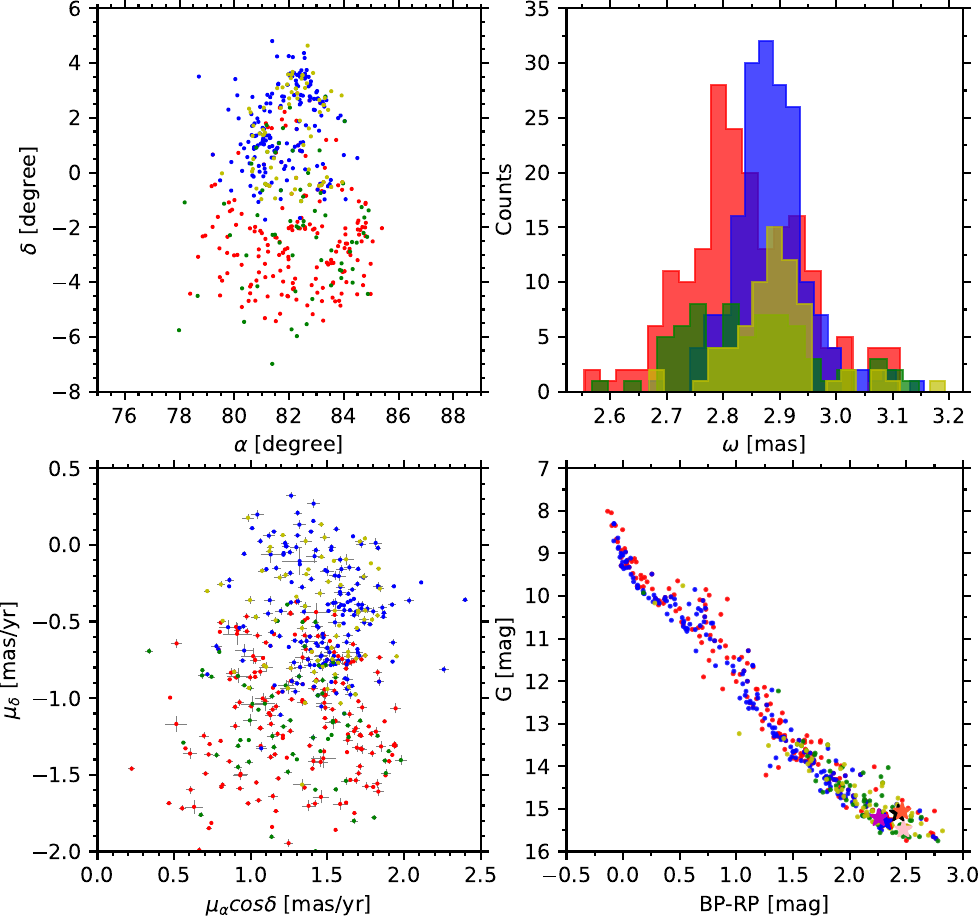}
	\caption{Distributions of the members of ASCC~19 and ASCC~21 with a radial velocity-to-error ratio greater than or equal to 3 in multi-dimensional parameter spaces. Members with normal radial velocities, corresponding to those within the red boxes in Fig.~\ref{Double_clusters_rv}, are marked by red (ASCC~19) and blue (ASCC~21), respectively, with members with abnormal radial velocities that are those outside of the red boxes in Fig.~\ref{Double_clusters_rv} being colored by green (ASCC~19) and yellow (ASCC~21). \textit{Top left panel}: Distribution of the member stars of both clusters in the 2D celestial coordinate system. Gray bars represent the errors of $\alpha$ and $\delta$. \textit{Top right panel}: Parallax histograms for the member stars for both clusters. \textit{Bottom left panel}: Distribution of the member stars of both clusters in proper motion space. Gray bars represent the errors of proper motion in both directions. \textit{Bottom right panel}: CMD of the two clusters. Filled pentagrams represent the member stars with radial velocities exceeding 100~km$\cdot$s$^{-1}$ in both clusters. All member data are sourced from \citet{van23}.}
	\label{Double_clusters_rv_phase}
\end{figure*}

The CMDs of ASCC~19 and ASCC~21 in Fig.~\ref{Double_clusters} indicate that they appear to have an identical age. To examine this, we plotted the detailed CMDs for these two OCs, as shown in Fig.~\ref{Double_clusters_CMD}. This was obtained by fitting isochrones to the two clusters using the \texttt{PARSEC} v1.2S \citep{bres12,chen14,chen15,tang14,mari17,past19}. The isochrone fitting was performed with a range of ages from 1~Myr to 100~Myr, in steps of log(age)~=~0.05, and with a fixed value of A$_{V}$~=~0. The Gaussian means of the metal abundances from various survey data were used as input parameters for the isochrone fitting. The fitting method is based on Eq.~2 of \citet{liu19}, which allowed us to determine the optimal isochrone for the clusters. We used a two times step size as the error of the best fitting age. It is evident that ASCC~19 and ASCC~21 share an identical age (log(age) = 6.95~$\pm$~0.05), implying that they may be a primordial cluster pair.

\begin{figure}[ht]
	\centering
	\resizebox{\hsize}{!}{\includegraphics{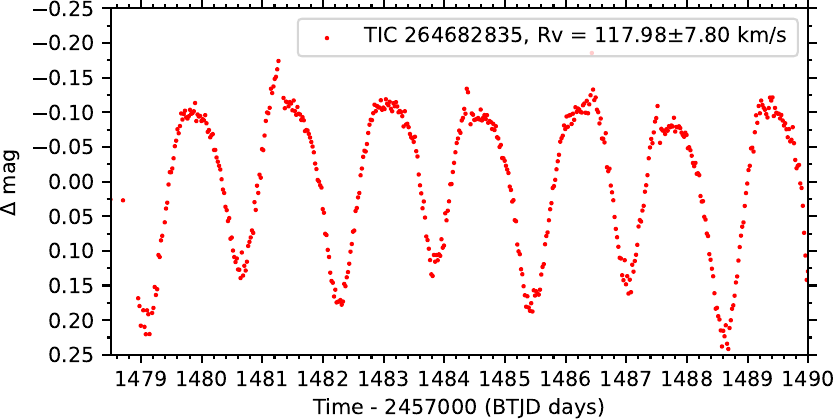}}
	\resizebox{\hsize}{!}{\includegraphics{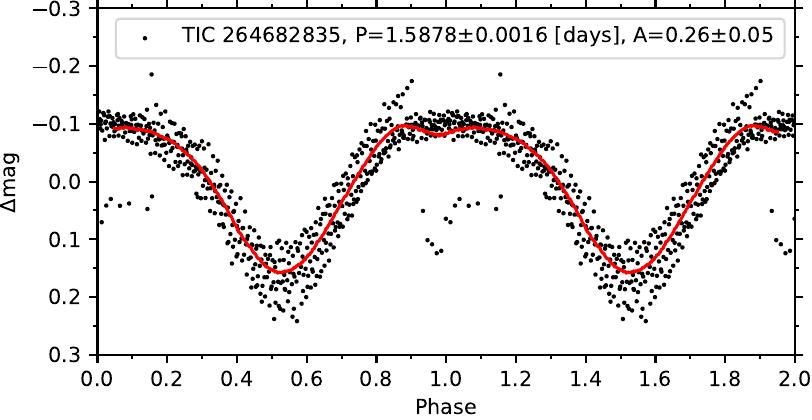}}
	\caption{\textit{Top panel}: Light curve of TIC~264682835. \textit{Bottom panel}:
		Phase diagram of \texttt{TIC~264682835}. The red curve represents the fitting curve to the phase. Its radial velocity, period, and amplitude are labeled in the diagram. All data are sourced from TESS.}
	\label{TIC_264682835}
\end{figure}

\subsection{Radial Velocity}

To further verify whether the cluster pair is primordial, it is important to examine their radial velocities (Rv) in addition to their ages. Accordingly, we plotted the histogram of the radial velocities for both clusters, as shown in Fig.~\ref{Double_clusters_rv}. To ensure the reliability of the radial velocity, we restricted the members with radial velocities that are at least three times greater than their measurement error.

\begin{figure}[ht]
	\centering
	\resizebox{\hsize}{!}{\includegraphics{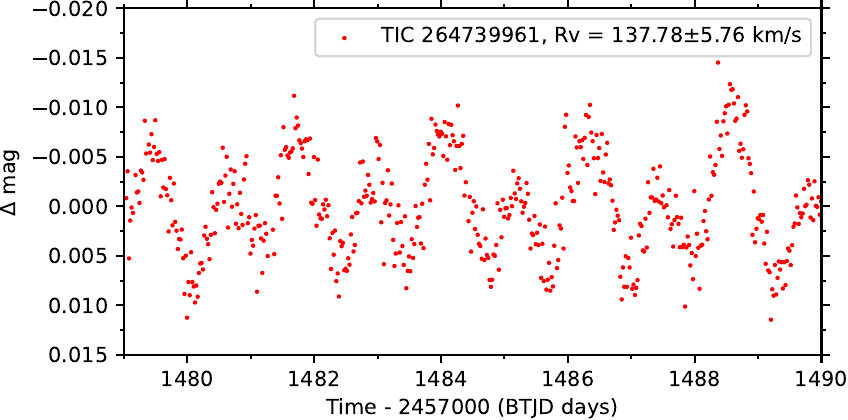}}
	\resizebox{\hsize}{!}{\includegraphics{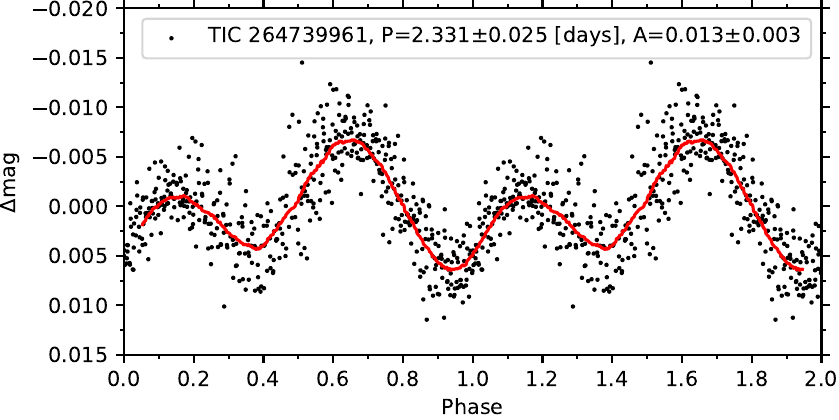}}
	\caption{\textit{Top panel}: Light curve of \texttt{TIC~264739961}. \textit{Bottom panel}: Phase diagram of \texttt{TIC~264739961}. All symbols are the same as those in Fig.~\ref{TIC_264682835}. All data are sourced from TESS.}
	\label{TIC_264739961}
\end{figure}

Figure.~\ref{Double_clusters_rv} displays the histograms of the radial velocities of the cluster pair. Based on this distribution, we calculated the median radial velocity and its median error for each cluster. The radial velocities of the two clusters are consistent within their error ranges, which aligns with the expected motion characteristics of a primordial cluster pair. Therefore, we concluded that ASCC~19 and ASCC~21 are likely a primordial cluster pair.

\subsection{Metallicity}

Metallicity is also one of the factors that verify whether our sample pair is primordial. Thus, we here analyzed the metallicity of the sample pair. We opted to acquire metal abundance data from alternative spectroscopic surveys through cross-matching. Consequently, we obtained metal abundance data for 57, 50, 144, and 119 member stars of ASCC~19 from the low-resolution spectrum (LRS) of the Large Sky Area Multi-Object Fiber Spectroscopic Telescope eleventh data release (LAMOST-DR11\footnote{\url{https://www.lamost.org/dr11/}} \citep{luo12, zhao12}), the medium-resolution spectrum (MRS) of LAMOST-DR11, the Apache Point Observatory Galactic Evolution Experiment seventeenth data release (APOGEE-DR17\footnote{\url{https://www.sdss4.org/dr17/irspec/spectro_data/}} \citep{abdu22}), and the fourth data release of the Galactic Archaeology with HERMES (GALAH-DR4\footnote{\url{https://www.galah-survey.org/dr4/overview/}} \citep{bude24}), respectively. Similarly, for ASCC~21, we acquired metal abundance data for 54, 40, 114, and 159 member stars from the same respective sources. The metal abundance values of these stars are at least three times their measurement error.

We then plotted the histograms of the metallicities for ASCC~19 and ASCC~21, as shown in Figs.~\ref{Double_Fe_H_four}. The systematic metallicities ([Fe/H]) for ASCC~19 were determined to be $-$0.16~$\pm$~0.22~dex, $-$0.14~$\pm$~0.25~dex, $-$0.12~$\pm$~0.06~dex, and $-$0.36~$\pm$~0.17~dex, while those for ASCC~21 were $-$0.15~$\pm$~0.21~dex, $-$0.12~$\pm$~0.04~dex, $-$0.12~$\pm$~0.07~dex, and $-$0.40~$\pm$~0.21~dex, all derived from Gaussian fits. Despite some inconsistencies in metal abundance across different surveys for each cluster, similar metal abundance within the same survey for the pair suggests they share a similar age. Therefore, we concluded that the cluster pair comprising ASCC~19 and ASCC~21 constitutes a primordial cluster pair.

\section{Properties of Primordial Cluster Pair}

\subsection{Possible Formation Process}

Fig.~\ref{Double_Fe_H_d} illustrates the distribution between the [Fe/H] metallicity parameters of the members, sourced from Gaia~DR3, APOGEE-DR17, and GALAH-DR4, and their distances to the clusters' centers. We noted that since the metal abundance parameter of \texttt{Gaia} is reliable for the FGK-type stars \citep{soub22, reci23, soub24}, here we exclusively analyzed the member stars with effective temperatures ranging from 3500~K to 7500~K. We found from Fig.~\ref{Double_Fe_H_d} that both clusters are broadly composed of member stars exhibiting two different metal abundances, as evidenced by the dispersion distribution on the left and the banded aggregate distribution on the right within the same panel (left or right) for each survey dataset (color-coded in green, or yellow, or blue). This suggests that the formation of these two clusters could be attributed to the amalgamation of several subclusters. Besides, this merger appears to be well-mixed and homogeneous, as the member stars with different metal abundances are evenly distributed throughout the clusters, extending from the central regions to the outer periphery, as shown in Fig.~\ref{Double_Fe_H_d}.

It is widely known that the fragmentation of the same giant molecular cloud \citep{mucc12} can lead to a certain number of small stellar groups or subclusters. Considering the similarities of ASCC~19 and ASCC~21 in ages, radial velocities, and metallicities, we thus inferred that the primordial cluster pair consisting of two clusters was likely formed through the fragmentation of the same molecular cloud.

\subsection{Mass Segregation}

We can see from Fig.~\ref{Double_clusters_rv} that a certain number of member stars are outside the red boxes. This implies that the radial velocities of these member stars are anomalous as opposed to the systematic radial velocities of their host clusters. It is widely accepted that these member stars with anomalous radial velocities may be due to two-body relaxation if their masses are small. For a cluster, the process results in less massive member stars gaining more kinetic energy and gradually moving to the outskirts of the cluster, while more massive stars lose kinetic energy and sink toward the cluster's center. Eventually, it could lead to a mass segregation phenomenon \citep[e.g.,][]{vesp09, port10, evan22, noor23}.

To investigate whether mass segregation exists in the primordial cluster pair, we employed the minimum spanning tree method defined by \citet{alli09} to quantitatively assess the mass segregation within the components of the pair. \citet{alli09} introduced the ``mass segregation ratio" (\texttt{$\Lambda_{MSR}$}), where a value greater than 1 indicates the existence of mass segregation. This method was also used by \citet{zhang20} to analyze the mass segregation in Blanco~1. The Python code for mass segregation used in this work is sourced from the \texttt{gaia\_oc\_amd} repository published by \citet{van23} on \texttt{GitHub}\footnote{\url{https://github.com/MGJvanGroeningen/gaia_oc_amd}}. Figure~\ref{mass_segregation} shows the $\Lambda_{MSR}$ of ASCC~19 and ASCC~21. We found that ASCC~19 exhibits a slight indication of mass segregation, while ASCC~21 shows no such signal. Therefore, there is no significant mass segregation for both clusters.

\subsection{Members with Anomalous Radial Velocities}

Due to the membership probability equal to 1 for all members in Fig.~\ref{Double_clusters_rv}, those with anomalous radial velocities are also cluster members. Stars with anomalous radial velocities may provide crucial insights into the evolution of their host \citep{hu22}. In this section, we thus explored the distributions of the members with anomalous radial velocities across various spaces.

Figure~\ref{Double_clusters_rv_phase} shows the distribution of member stars with a radial velocity-to-error ratio greater than or equal to 3 across four different parameter spaces. We observe overlaps of these members in the position, parallax, proper motion, and CMD spaces. In addition, the distributions (green and yellow regions) of members with anomalous radial velocities extend throughout the internal structure of the cluster pair in parallax space. Meanwhile, these members are almost always faint stars, which can be inferred as less massive, as depicted in the bottom right panel of Fig.~\ref{Double_clusters_rv_phase}. Therefore, combined with the result of no mass segregation in the primordial cluster pair, we deduced that the clusters are beginning to undergo the two-body relaxation process, possibly at an early stage.

\begin{figure}[ht]
	\centering
	\resizebox{\hsize}{!}{\includegraphics{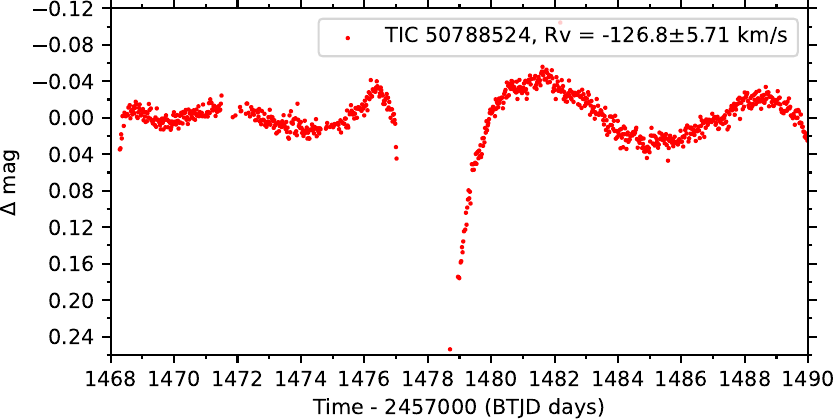}}
	\resizebox{\hsize}{!}{\includegraphics{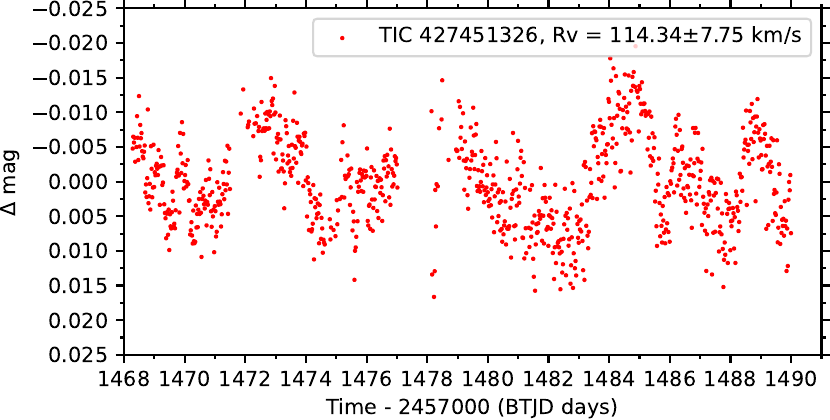}}
	\resizebox{\hsize}{!}{\includegraphics{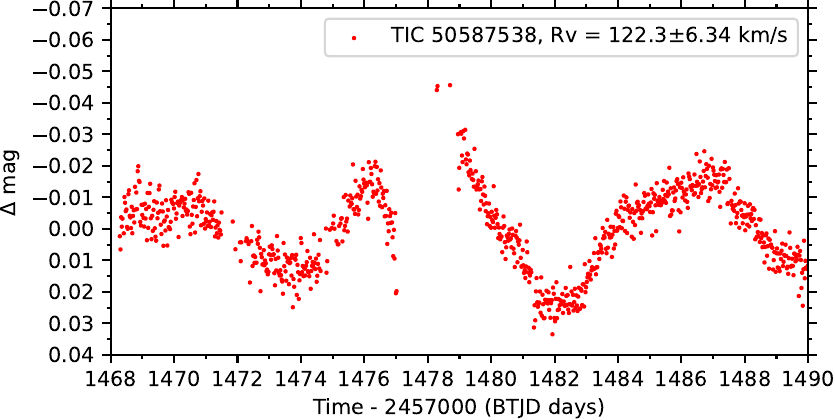}}
	\caption{\textit{Top panel}: Light curve of \texttt{TIC~50788524}. \textit{Middle panel}: Light curve of \texttt{TIC~427451326}. \textit{Bottom panel}: Light curve of \texttt{TIC~50587538}. All data are sourced from TESS.}
	\label{Three_ligthcurve}
\end{figure}

\begin{table*}[ht]
	\centering
	\caption{The parameters of the members with radial velocities exceeding 100~km$\cdot$s$^{-1}$}
	\label{table:members}
	\small
	\resizebox{\textwidth}{!}
	{\begin{tabular}{ccccccccc}
			\hline\noalign{\smallskip}
			\hline\noalign{\smallskip}
			(1) & (2) & (3) & (4) & (5) & (6) & (7) & (8) & (9)\\
			Gaia~DR3 ID & Ra & Dec & Parallax & Rv & $[$Fe/H$]$ & TIC ID & Host cluster & Symbols in Fig.~\ref{Double_clusters_rv_phase}\\
			-- & (degree) & (degree) & (mas) & (km$\cdot$s$^{-1}$) & (dex) & -- & -- & --\\
			\hline\noalign{\smallskip}
			3223872069503741568  & 82.30 & 2.89 & 2.912 & 117.98~$\pm$~7.80 & -- & 264682835 & ASCC~19 and ASCC~21 & Blue-filled pentagram\\
			3223841450681464448 & 82.43 & 2.55 & 2.948 & 137.78~$\pm$~5.76 & $-$0.133~$\pm$~0.056 & 264739961 & ASCC~21 & Coral-filled pentagram\\
			3220751346266572672  & 83.14 & $-$0.41 & 2.842 & $-$126.77~$\pm$~5.71 & $-$0.359~$\pm$~0.070 & 50788524 & ASCC~19 and ASCC~21 & Red-filled pentagram\\
			3217579220861913472  & 84.04 & $-$1.36 & 2.835 & 114.34~$\pm$~7.75 & $-$0.081~$\pm$~0.011 & 427451326 & ASCC~19 & Purple-filled pentagram\\
			3210550073087799296 & 82.17 & $-$4.22 & 2.928 & 122.30~$\pm$~6.34 & -- & 50587538 & ASCC~19 & Pink-filled pentagram\\
			3234192596742088192  & 80.65 & 1.70 & 2.934 & 107.20~$\pm$~6.97 & $-$0.246~$\pm$~0.065 & -- & ASCC~21 & Black-filled pentagram\\
			
			\hline\noalign{\smallskip}
	\end{tabular}}
	
\end{table*}

If a cluster is in the early stages of the two-body relaxation process, the number of its member stars with large velocities may be relatively small or even absent because the energies of the member stars have not yet reached a state of equalization. To explore it, we conducted a detailed analysis of the member stars with radial velocities exceeding 100~km·s$^{-1}$ (denoted by filled pentagrams in the bottom right panel of Fig.~\ref{Double_clusters_rv_phase}). We identified six members with radial velocity greater than 100~km·s$^{-1}$ in the cluster pair. Among these, four belong to ASCC~19, four ASCC~21, and two to both ASCC~19 and ASCC~21. The parameters of these six stars are presented in Table~\ref{table:members}. It is well known that the light variation of stars can influence the measurement of their intrinsic radial velocities. Therefore, It is crucial to verify whether these stars are variable.

We successfully retrieved and downloaded the photometric data for five of the six members with anomalous radial velocities we identified from the Transiting Exoplanet Survey Satellite (TESS) database\footnote{\url{https://archive.stsci.edu/missions-and-data/tess}} \citep{rick14}. To determine whether these stars are variable, we plotted their lightcurves and phase diagrams, using the \texttt{Python} package \texttt{LightKurve}\footnote{\url{https://github.com/lightkurve/lightkurve}} \citep{ligh18} and the \texttt{PERIOD04} software \citep{lenz05}. Figure~\ref{TIC_264682835} shows the light curve and phase of \texttt{TIC~264682835}, a member with a radial velocity of 117.98~$\pm$~7.80~km·s$^{-1}$. This star appears to be a periodic variable, previously classified as BY Draconis-type variable (also known as a rotational variable) \citep{chen20}. \citet{chen20} noted that the rotational periods of BY Draconis-type variables range from 0.25 to 20 days. We found that the period of \texttt{TIC~264682835} is approximately 1.5878~$\pm$~0.0016~days, which falls within this range, as shown in Fig.~\ref{TIC_264682835}. Additionally, this star is located at the lower edge of the main sequence in CMD of Fig.~\ref{Double_clusters_rv_phase}. \citet{tsan22} reported its radial velocity as Rv = 20.418~$\pm$~0.921~km·s$^{-1}$ based on data from APOGEE. Furthermore, \citet{verb24} provided its Rv of 302.009~$\pm$~290.928~km·s$^{-1}$ based on Gaia BP/RP spectra. Since the star's radial velocity varies across different epochs, combined with its observed light variation, we inferred that it is a variable star. This suggests that its radial velocity may not represent its intrinsic radial velocity.

We also plotted the light curve and phase diagram for \texttt{TIC~264739961}, which has a radial velocity (Rv) of 137.78~$\pm$~5.76~km·s$^{-1}$, as shown in Fig.~\ref{TIC_264739961}. We found that this star is also a periodic variable, exhibiting a period of 2.331~$\pm$~0.025~days and an amplitude of 0.013~$\pm$~0.003~mag. \citet{sern21} classified it as a young star with a projected rotational velocity of 23.3~$\pm$~0.8~km·s$^{-1}$ based on APOGEE data. \citet{abdu22} reported a radial velocity of Rv = 39.20~$\pm$~0.08~km·s$^{-1}$ for this star, based on the seventeenth data release of the Sloan Digital Sky Surveys (SDSS-DR17). Additionally, other reports show its radial velocity as Rv = 24.1~km·s$^{-1}$ \citep{ding22}, Rv = 19.3~$\pm$~9.7~km·s$^{-1}$ \citep{zhan21} based on LAMOST data, and Rv = 20.255~$\pm$~0.355~km·s$^{-1}$ \citep{tsan22} based on APOGEE data. Therefore, we concluded that the radial velocity presented in this work is not its intrinsic radial velocity.

Furthermore, Fig.~\ref{Three_ligthcurve} presents the light curves for three additional stars with anomalous radial velocities, \texttt{TIC~50788524}, \texttt{TIC~427451326}, and \texttt{TIC~50587538}. The photometric variations observed in these light curves suggest that these stars are variables. For \texttt{TIC~50788524}, \citet{verb24} reported a radial velocity of 110.501~$\pm$~146.581~km·s$^{-1}$. \citet{abdu22} also provided several radial velocities based on APOGEE data: 31.946~$\pm$~0.049~km·s$^{-1}$, 18.038~$\pm$~0.067~km·s$^{-1}$, and 20.059~$\pm$~0.065~km·s$^{-1}$. Similarly, \texttt{TIC~427451326} was studied by \citet{abdu22}, who reported Rv values of $-$5.27~$\pm$~0.008~km·s$^{-1}$, 43.74~$\pm$~0.06~km·s$^{-1}$, $-$4.03~$\pm$~0.05~km·s$^{-1}$, $-$5.61~$\pm$~0.06~km·s$^{-1}$, and 43.63~$\pm$~0.07~km·s$^{-1}$. \citet{spin21}, \citet{swig21}, and \citet{tsan22} published Rv values of 19.98~km·s$^{-1}$, 20.11~$\pm$~0.02~km·s$^{-1}$, and 19.97~$\pm$~0.30~km·s$^{-1}$, respectively. In addition, \citet{verb24} reported a Rv of $-$118.73~$\pm$~158.95~km·s$^{-1}$ for \texttt{TIC~50587538}. Finally, for the one without TESS data (denoted by the black-filled pentagram in Fig.~\ref{Double_clusters_rv_phase}), \citet{hern23} identified it as a T Tauri star, which is generally a young stellar object candidate. Based on all the radial velocity parameters mentioned above, we can conclude that the radial velocities of these four stars are likely not their intrinsic radial velocities.

Considering the radial velocities of these stars as reported in the aforementioned literature, along with their ages and positions derived from the CMD, we can infer that these six stars are variables, which may be not only possibly individual young variables, but also binary. This is because the changes in radial velocities of very young variable stars (e.g. Classical T Tauri stars) are often attributed to magnetospheric accretion and the presence of hotspots on the stellar surface during observation \citep{petr11, petr23}. Besides, these young variables are likely protostars based on ages in the CMD. They are likely in binary or higher multiplicity systems with large gas or debris disks \citep{jorg22}.

Ultimately, we concluded that the members with anomalous radial velocities exceeding 100~km·s$^{-1}$ do not show correspondingly large intrinsic radial velocities. In other words, their large radial velocities observed in this study cannot be attributed to energy equalization. This suggests that the primordial cluster pair is in the early stages of two-body relaxation.

\begin{figure*}[ht]
	\centering
	\includegraphics[width=110mm]{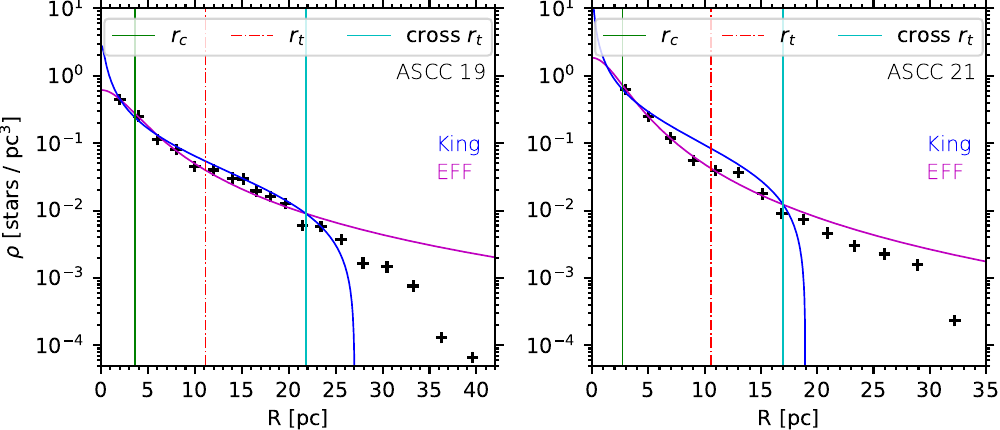}
	\caption{Radial density profiles of ASCC~19 (left) and ASCC~21 (right) in 3D spatial space. The stellar radial densities in each subplot are represented by black crosses. The purple and blue curves correspond to the \texttt{EFF} and \texttt{King} models, respectively. The green and cyan solid lines denote the core radius estimated by the \texttt{EFF} model and the tidal radius obtained from the intersection of the two density profile fitting lines, respectively. The red dotted lines represent the tidal radii calculated using the \texttt{von Hoerner} formula in Sec.~\ref{Sec: discussion}. The parameters presented in this figure are calculated via the member data sourced from \citet{van23}.}
	\label{radial_density_profile}
\end{figure*}

\begin{figure*}[ht]
	\centering
	\includegraphics[width=150mm]{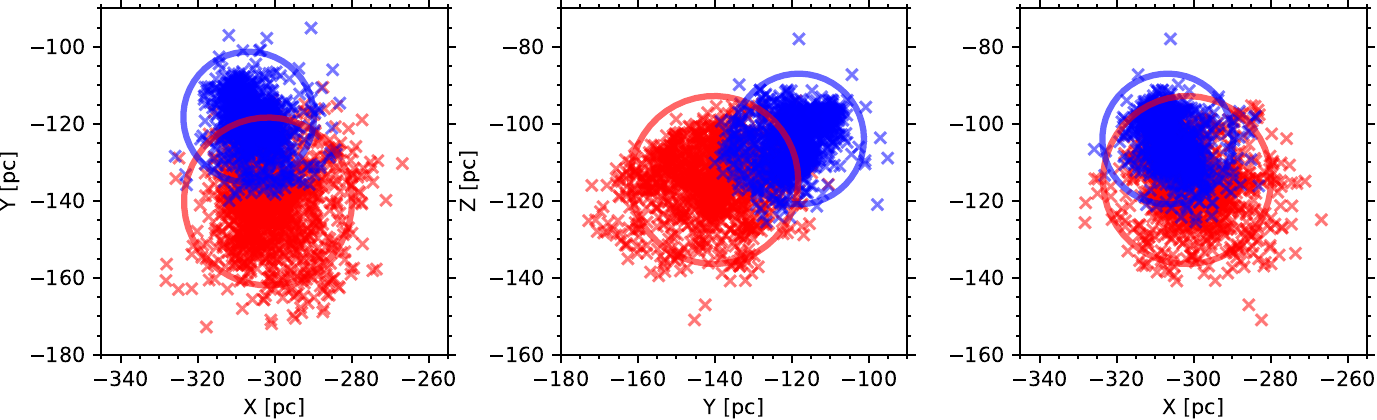}
	\caption{The 3D projection distribution of the cluster pair on X-Y (left), Y-Z (middle), and X-Z (right) planes. The red and blue crosses indicate the members of ASCC~19 and ASCC~21, respectively. The red and blue circles indicate the tidal radii, derived from the intersection points, as detailed in Fig.~\ref{radial_density_profile}. The parameters presented in this figure are calculated via the member data sourced from \citet{van23}.}
	\label{ASCC1921_XYZ}
\end{figure*}

\subsection{3D Spatial Morphology}\label{morphology}

The study of the morphology of OCs can provide observational insights into their formation and evolution \citep{hu21a, hu21b}. Therefore, in this section, we assessed the dynamical state of the primordial cluster pair by investigating its morphology.

\begin{figure}[ht]
	\centering
	\includegraphics[width=80mm]{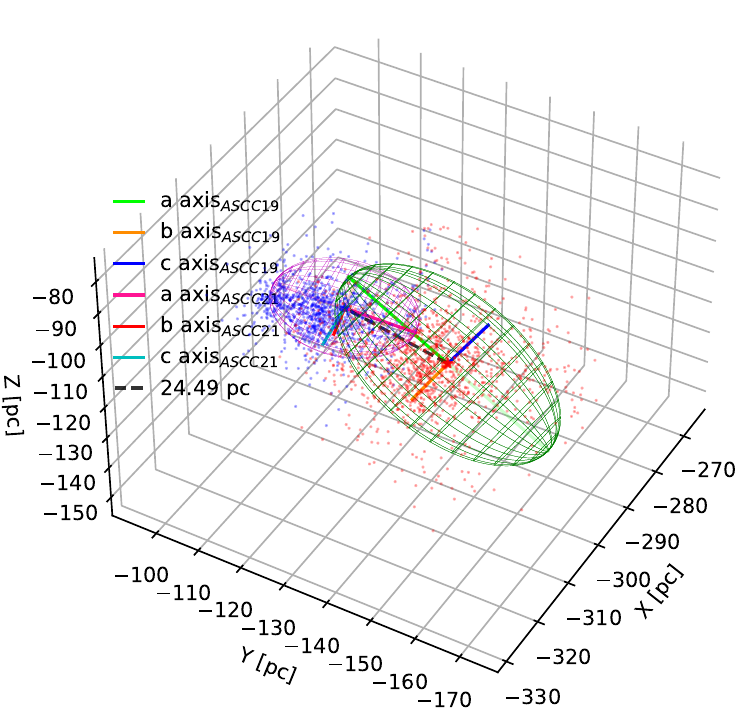}
	\caption{3D spatial structure of the cluster pair (ASCC~19 and ASCC~21), shown with ellipsoidal curves. The green ellipsoid, with its fitted center marked by a red-filled pentagram, represents ASCC~19, while the purple ellipsoid, with its fitted center marked by a blue-filled pentagram, represents ASCC~21. The small red and blue dots represent the member stars of ASCC~19 and ASCC~21, respectively. The black dashed line indicates the distance between the fitted centers of the two clusters, which is 24.49~pc. The differently colored bars represent the various axes of the ellipsoids, as shown in the legend. The parameters presented in this figure are calculated via the member data sourced from \citet{van23}.}
	\label{Double_clusters_shape}
\end{figure}

Tidal radii and the 3D morphology of clusters were combined to diagnose the dynamical state of the primordial cluster pair. We adopted the \texttt{EFF} model \citep{elso87} and the \texttt{King} model \citep{king62} to fit their radial density profiles in 3D space to obtain the tidal radii. The stellar radial density profile of a typical OC is generally composed of three components: the core, the bulk, and the tidal debris. Our target clusters align with this assumption. However, \citet{king62} and \citet{elso87} focused on the reproduction of the cluster's bulk and external tidal structure via the \texttt{EFF} and \texttt{King} models, respectively. It would be expected that a single model would not suffice for an accurate radial density profile fit for our clusters. Therefore, we combined the two models to derive the tidal radii of ASCC~19 and ASCC~21.

Below are the formulas of the \texttt{EFF} and \texttt{King} models utilized in our study:

For the \texttt{EFF} model:

\begin{equation}\label{EFF}
	\rho (r) = \rho _{0} (1+(\frac{r}{r_{c}})^{2})^{\frac{-\eta}{2} },
\end{equation}

where $\rho_{0}$ and $r_{c}$ denote the central density of clusters and the core radius, respectively. $\eta$ is the slope of the \texttt{EFF} template for $r$ much larger than $r_{c}$.

For the \texttt{King} model:

\begin{equation}\label{King}
	\centering
	\rho (r) = \rho_{0} \cdot (\frac{1}{\sqrt{1+(r/r_{c} )^{2} } } - \frac{1}{\sqrt{1+(r_{t}/r_{c} )^{2}}} )^{2} + \rho_{bg},
\end{equation}

where $\rho_{0}$ and $r_c$ in the \texttt{King} model have the same meaning as in the \texttt{EFF} template. $r_{c}$ is also the distance from the clusters' density center to the point where $\rho (r)$ = $\rho_{0}$/2, with $r_{t}$ being the tidal radius and also the point where $\rho (r)$ = $\rho_{bg}$. $\rho_{bg}$ is the background density.

The fitting process is as follows: First, we calculated the distances of the members to the center of their respective host clusters, using their coordinates (X, Y, and Z). Second, the 3D spatial region encircling the clusters' center was partitioned into some spherical bins sharing the same central point. Third, we recorded the radial distance of each bin and computed the stellar number density within those bins. Finally, these two sets of parameters were inputted into the \texttt{EFF} and \texttt{King} models to derive the radial density profile curves, as illustrated in Fig.~\ref{radial_density_profile}.

Figure~\ref{radial_density_profile} shows the stellar radial density profiles of our two clusters in 3D spatial space along with the fitting lines derived from the \texttt{EFF} and \texttt{King} models. It is evident that the \texttt{EFF} provides a better fit for both clusters within the range of $R$~$\leq$~20~pc, compared to outside this range. The chi-square values obtained from the \texttt{King} model for these clusters are 7.07 for ASCC~19 and 36.61 for ASCC~21, respectively, while the chi-square values fitted by the \texttt{EFF} are 0.54 for ASCC~19 and 0.16 for ASCC~21. Therefore, we adopted the core radii determined by the \texttt{EFF} model for our sample clusters, with $r_{c}$~=~3.59~$\pm$~0.31~pc for ASCC~19 and $r_{c}$~=~2.73~$\pm$~0.29~pc for ASCC~21. Additionally, the core radii of ASCC~19 and ASCC~21 calculated by \citet{hunt23} are $r_{c}$~=~4.66~pc and $r_{c}$~=~2.55~pc, respectively, which are nearly consistent with our results.

It is noted that the tidal radii fitted by the \texttt{King} template were not taken as the final tidal radii for our sample clusters due to its poor performance across the entire range of fits. However, we can consider the point where the radial density profiles fitted by these two models intersect as the tidal radii of the clusters, with $r_{t}$ being about 21.82~pc for ASCC~19 and 16.97~pc for ASCC~21. The tidal radii of the two clusters have also been estimated in \citet{hunt23}, yielding $r_{t}$~=~4.66~pc for ASCC~19 and $r_{t}$~=~10.63~pc for ASCC~21, respectively. This shows that our tidal radii are much larger than theirs, which may be attributed to the greater number of cluster members in the present study. Moreover, the parameters from \citet{hunt24} show that the average tidal radius of OCs at the Solar Circle ($\leq$~500~pc) is nearly 19~pc, which means the tidal radii of our sample are more likely accurate. The core and tidal radii parameters we obtained are also presented in Table~\ref{table:data}.

\begin{figure*}[ht]
	\centering
	\includegraphics[width=150mm]{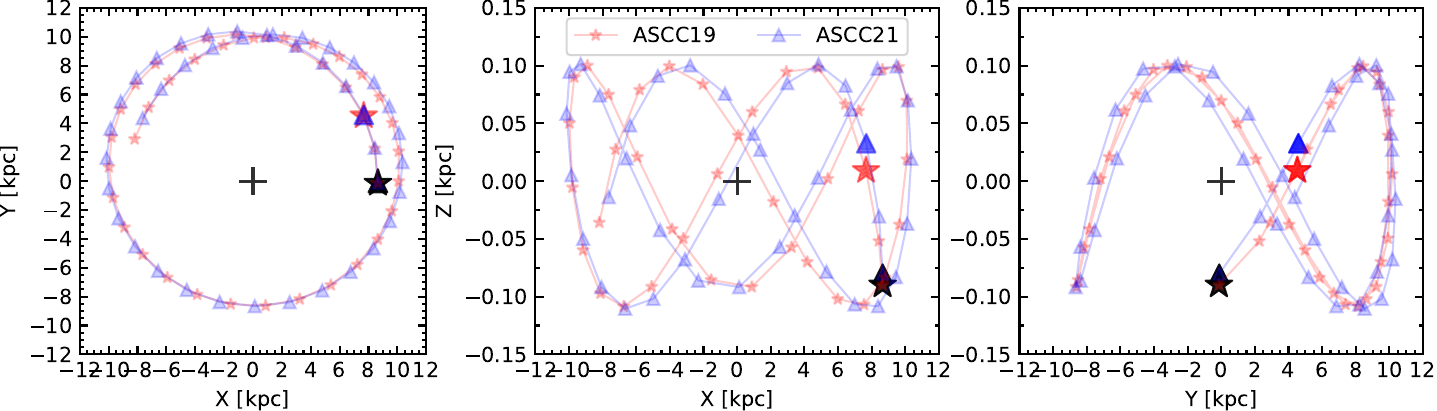}
	\caption{The 3D projected motion trajectories of ASCC~19 and ASCC~21 in the X-Y (left), Y-Z (middle), and X-Z (right) planes. The current positions of ASCC~19 and ASCC~21 are marked by a black-filled pentagram and a black-filled triangle, respectively. The positions after 40~Myr are denoted by a red-filled pentagram (ASCC~19) and a blue-filled triangle (ASCC~21). The time span between two neighboring symbol marker points is 20~Myr for both red and blue trajectories. The parameters presented in this figure are calculated via the member data sourced from \citet{van23}.}
	\label{Triple_galpy_XYZ}
\end{figure*}

Next, we proceeded to determine the 3D projected distribution of the members of the cluster pair. We thus firstly calculated their 3D spatial coordinates within the heliocentric Cartesian coordinate system (X, Y, and Z) \footnote{The reference frame is an XYZ Cartesian coordinate system, with the Sun at its center. The positive X-axis is oriented from the Sun's projected position on the Galactic midplane towards the Galactic center, with the positive Y-axis aligning with the direction of the Galactic rotation. The positive Z-axis extends toward the north pole of the Galaxy.}. The coordinate calculation for each member of the pair was facilitated by the \texttt{Python} \texttt{Astropy} package \citep{astr13, astr18}. However, because \texttt{Gaia}'s parallax has a symmetric error bar, direct inversion of the parallax can result in pseudo-stretching of the cluster's 3D morphology along the line-of-sight direction \citep[see, e.g.,][]{bail15, luri18, carr19}, we adopted a Bayesian distance correction model developed by \citet{carr19} and \citet{pang21} to mitigate this problem, which has been employed in some studies \citep[e.g.,][]{ye21, qin23, hu23, hu24}. Eventually, we obtained the corrected distances of the members of ASCC~19 and ASCC~21 and further estimated their 3D spatial coordinates. Figure~\ref{ASCC1921_XYZ} shows the 3D projected morphologies of the cluster pair on the X-Y, Y-Z, and X-Z planes. We found an overlap in the 3D projected distribution of the members of the two clusters, as well as their tidal circles. Therefore, we speculated that there is a gravitational interaction between the clusters.

Subsequently, we proceeded to fit a 3D ellipsoid to the 3D spatial distribution of the member stars of ASCC~19 and ASCC~21, a method that has also been employed by \citet{hu24}. Figure~\ref{Double_clusters_shape} displays the 3D spatial structure of the primordial cluster pair, along with the 3D fitting ellipsoids of its components. Similar to the result in Fig.~\ref{ASCC1921_XYZ}, there is also a crossover in their 3D fitting ellipsoids. Furthermore, we determined that the distance between the centers of the fitted ellipsoids of ASCC~19 and ASCC~21 is approximately 24.49~pc, which exceeds half the sum of their tidal radii obtained with the intersection point.

In order to verify whether the distance between the centers is correct, we calculated the 3D separation of the clusters and compared it with the distance between the centers. The 3D separation of the clusters can be estimated as $a$~=~27.00~$\pm$~7.51~pc, based on the clusters' coordinates (X, Y, Z) listed in Table~\ref{table:data}. This calculation was performed by the Monte Carlo method, with 10000 iterations sampling for X, Y, and Z, using their respective errors as the sampling variances and these coordinates as the means. Obviously, the centers' distance is in agreement with the separation we determined. In light of the above, it also indicates the cluster pair is likely undergoing a mutual tidal interaction process.

\subsection{Motion Trajectories}

We have determined that a mutual tidal interaction is occurring between ASCC~19 and ASCC~21, suggesting the potential for these two clusters to eventually merge into a single object. To explore this further, we simulated the motion trajectories of both clusters, using the \texttt{galpy.potential} module (\texttt{MWPotential2014}) within the \texttt{Python} \texttt{galpy} \citep{bovy15} package. It is an axisymmetric potential module consisting of a Miyamoto-Nagai disc \citep{miya75}, a bulge, and a dark matter halo modeled with a Navarro-Frenk-White (NFW) potential \citep{nava97}. By inputting the position, parallax, proper motion, and radial velocity parameters of ASCC~19 and ASCC~21 into the module, we integrated the potential to derive the 3D spatial motion trajectories of these two clusters over time.

Figure~\ref{Triple_galpy_XYZ} presents the projected motion trajectories of the two clusters on the X-Y (left), Y-Z (middle), and X-Z (right) planes within 400~Myr forward. It is evident that the two clusters complete approximately one and a half orbits around the galactic center within this time frame. The current positions of ASCC~19 and ASCC~21 are marked by a black-filled pentagram (for ASCC~19) and a black-filled triangle (for ASCC~21) within the galactic disk. After 40~Myr, they will appear in the positions denoted by a red-filled pentagram (for ASCC~19) and a blue-filled triangle (for ASCC~21), as shown in Fig.~\ref{Triple_galpy_XYZ}. Based on these motion trajectories, it appears unlikely that ASCC~19 and ASCC~21 will merge into a single cluster over time.

\section{Discussion}\label{Sec: discussion}

In this section, we aimed to figure out the nature of the primordial cluster pair. Does it constitute a physical cluster pair, i.e., a binary cluster? \citet{minn04} pointed out that the relevant size of a binary source is the Roche radius, which is defined by the following formula:

\begin{equation}
	R_{R} = a(0.38+0.2\log\frac{m_{1}}{m_{2}} )^{\frac{1}{2}},
\end{equation}

where $a$ is the semimajor orbital axis and also the separation between two objects \citep{minn04}, with $m_{1}$ and $m_{2}$ being the individual cluster masses \citep{pacz71}. The mass parameters of our sample clusters are given in Appendix~\ref{appendix} and also listed in Table~\ref{table:data}. According to this formula, the Roche radius of the clusters can be calculated as $R_{R}$~$\approx$~16.88~$\pm$~4.70~pc in the case of our sample $a$~=~27.00~$\pm$~7.51~pc.

Furthermore, \citet{kim00} provided an equation to estimate the tidal radius of a cluster embedded within the potential of a massive galaxy, which could be compared with the Roche radius above \citep{minn04}. The formula, akin to the von Hoerner's equation \citep{von57, minn04}, is as follows:

\begin{equation} \label{Rt}
	R_{t} = (\frac{M_{c}}{2M_{G}})^{1/3}  \times R_{G},
\end{equation}

where $M_{c}$ and $M_{G}$ are the cluster's mass and the enclosed mass of the galaxy within $R_{G}$ that is the distance to the galaxy center, respectively \citep{von57}. The $R_{G}$ of our clusters are presented in Table~\ref{table:data}. The Galactic enclosed mass within $R_{G}$ for a cluster is given by \citet{genz87}:

\begin{equation}
	M_{G} = 2 \times 10^{8}\text{M}_{\odot}(\frac{R_{G}}{30\text{pc}})^{1.2}.
\end{equation}

Thus, the estimated tidal radii turn out to be $R_{t}$~=~11.12~$\pm$~0.003~pc for ASCC~19 and $R_{t}$~=~10.58~$\pm$~0.002~pc for ASCC~21, as indicated by the red dotted lines in Fig.~\ref{radial_density_profile}. It is apparent that the tidal radii derived from the Eq.~\ref{Rt} are smaller than the Roche radius of our sample. Therefore, our sample should be an unbound cluster pair, that is, a double cluster.

Additionally, to test the conclusion above, we employed the velocity criterion provided by \citet{van98} and \citet{mucc12}, that it is to compare the observational differences of clusters' velocities with the maximum expected differences of clusters' velocities. The difference between the average radial velocity values of our clusters is $\Delta V_{Rv}$~=~1.29~km·s$^{-1}$. To calculate the maximum expected velocity difference, we first determined the orbital period of our sample based on Kepler's third law, which yielded a value of $P_{orb}$~$\approx$~352~$\pm$~143~Myr for separation distance $a$~=~27.00~$\pm$~7.51~pc. Then the orbital velocity of the sample clusters orbiting each other was calculated to be $V_{orb}$~$\approx$~0.503~$\pm$~0.084~km·s$^{-1}$, via the formula $V_{orb}$ = 2$\pi$$a$/$P_{orb}$. This value represents the maximum expected difference of the two clusters' velocities \citep{mucc12,mora19}, but which is still smaller than their velocities' difference in observations. Therefore, it supports our conclusion that the two clusters form an unbound pair.

\section{Summary}

Based on the literature data, we for the first time discovered and confirmed that ASCC~19 and ASCC~21 constitute a new primordial cluster pair, as evidenced by their ages, kinematics, and metallicities. Keys findings about this primordial cluster pair are as follows:

1. The similarities in position, parallax, proper motion, CMD, radial velocity, and metallicity indicate that the primordial cluster pair was likely born from the fragmentation of the same molecular cloud.

2. A detailed analysis of the distribution between the metal abundances and distances to clusters' centers suggests that the formation of the two clusters likely resulted from the merger of multiple subclusters.

3. The primordial cluster pair shows no significant mass segregation and contains a certain number of members with anomalous radial velocities, suggesting that the clusters are in the early stage of the two-body relaxation.

4. By analyzing photometric data from TESS, combining it with radial velocity parameters from the literature, and examining the members' positions in the CMD, we found the stars with anomalous radial velocities exceeding 100~km·s$^{-1}$ are young variables. Thus, their radial velocities are likely not intrinsic.

5. We diagnosed the dynamical state of the primordial cluster pair by analyzing its 3D projected distribution on the X-Y, X-Z, and Y-Z planes, as well as its 3D morphological structure. The results indicate that the pair is undergoing a mutual tidal interaction process.

6. The Roche radius of the primordial cluster pair exceeds the tidal radii of its components, suggesting that it is not a bound system. This conclusion is further verified via the comparison of the radial velocity difference between the two clusters and their maximum expected velocity difference. Thus, the primordial cluster pair is a double cluster.

7. Orbital simulations also show that the clusters will not merge but will continue along similar orbits.


\acknowledgments
We would like to warmly thank the anonymous reviewer for suggesting improvements to the manuscript. This work is supported by the National Natural Science Foundation of China (NSFC) under grant 12303037, the Fundamental Research Funds of China West Normal University (CWNU, No.493065), the National Key R\&D Program of China (Nos. 2021YFA1600401 and 2021YFA1600400), the National Natural Science Foundation of China (NSFC) under grant 12173028, the Chinese Space Station Telescope project: CMS-CSST-2021-A10, the Sichuan Youth Science and Technology Innovation Research Team (Grant No. 21CXTD0038), and the Innovation Team Funds of China West Normal (No. KCXTD2022-6). Qingshun Hu would like to acknowledge the financial support provided by the China Scholarship Council program (Grant No. 202308510136). This study has made use of the Gaia DR3, operated by the European Space Agency (ESA) space mission (Gaia). The Gaia archive website is \url{https://archives.esac.esa.int/gaia/}. All TESS data presented in this paper can be found in the Mikulski Archive for Space Telescopes (MAST) at the Space Telescope Science Institute (STScI): \dataset[10.17909/fhgz-ft26]{https://dx.doi.org/10.17909/fhgz-ft26}. STScI is operated by the Association of Universities for Research in Astronomy, Inc., under NASA contract NAS5-26555. Some of the radial velocities this work adopted are based on the data acquired through the Guoshoujing Telescope. Guoshoujing Telescope (the Large Sky Area Multi-Object Fiber Spectroscopic Telescope; LAMOST) is a National Major Scientific Project built by the Chinese Academy of Sciences. LAMOST is operated and managed by the National Astronomical Observatories, Chinese Academy of Sciences. This work also adopted the data from the Galactic Archaeology with HERMES (GALAH) survey. Software: \texttt{Astropy} \citep{astr18}, \texttt{Galpy} \citep{bovy15}, \texttt{LightKurve} \citep{ligh18}, and \texttt{TOPCAT} \citep{tayl05}.



\begin{appendix}

\section{Completeness estimation}\label{appendix}

\begin{figure}[ht]
	\centering
	\includegraphics[width=100mm]{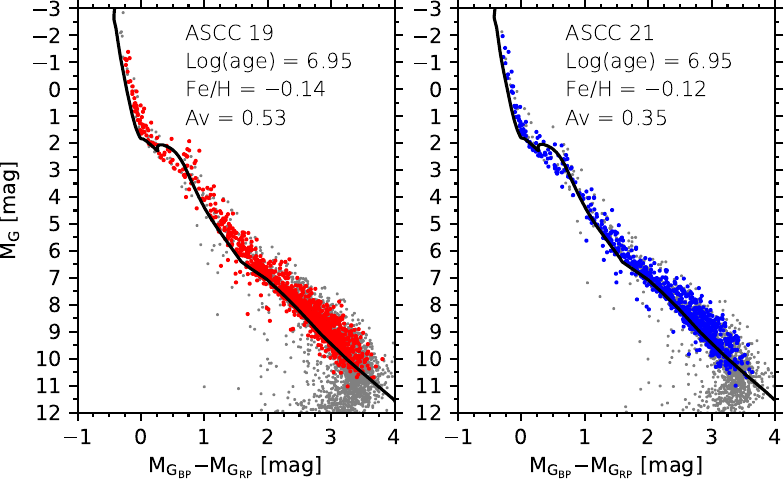}
	\caption{Same as Fig.~\ref{Double_clusters_CMD}. But all gray dots in each panel represent the cluster members with membership probability ranging from 0 to 1. All member data are sourced from \citet{van23}.}
	\label{CMD_comparison}
\end{figure}

In the process of filtering member stars, we found that the member stars of ASCC~19 and ASCC~21 are obviously mixed with field stars, as shown in Fig.~\ref{CMD_comparison}. Figure~\ref{CMD_comparison} illustrates the CMD distributions of ASCC~19 and ASCC~21 before and after applying the membership probability cut. The gray and colored dots distributions represent the CMD distributions of the two clusters before and after applying the cut, respectively. It is clear that not all the member stars marked by gray dots are centrally distributed in the main sequence belt, which means that the member stars far away from the main sequence belt are most likely to be caused by field star contamination or by extinction \citep{cao24}. Therefore, we restrict the probability of member stars, equal to 1 so as to minimize the affection caused by these two problems.

\begin{figure}[ht]
	\centering
	\includegraphics[width=110mm]{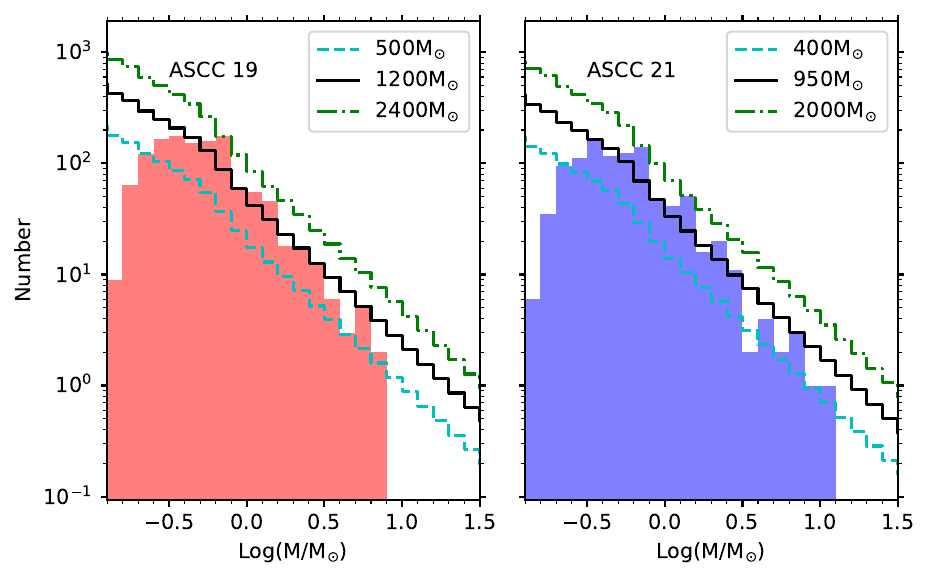}
	\caption{Initial stellar mass estimation of ASCC 19 and ASCC 21 after applying the membership probability cut. \textit{Left panel}: The observed mass distribution of ASCC 19, with fitted IMFs of varied total mass; \textit{Right panel}: The observed mass distribution of ASCC 21, with fitted IMFs of varied total mass. The parameters presented in this figure are based on the member data sourced from \citet{van23}.}
	\label{imf}
\end{figure}

The red and blue dots in Fig.~\ref{CMD_comparison} represent the member stars of ASCC~19 and ASCC~21 with member probability 1, respectively. Selecting reliable member stars based on membership probabilities may result in the exclusion of some true member stars. To explore this, we employ a fitted initial mass function (IMF) from \citet{krou01} to estimate the completeness of the selected member stars. What we used in the present work is the multiple part power-law IMF, similar to equation~6 of \citet{krou01}. Its lower and upper limits are set to 0.08~M$_{\odot}$ and 50~M$_{\odot}$, respectively. We constructed mass distribution histograms for the members of ASCC 19 and ASCC 21 and employed the IMF to generate a series of mass histogram profiles with varying initial masses, as shown in Fig.~\ref{imf}. The initial stellar mass for ASCC~19 is approximately 1200~M$_{\odot}$, while ASCC~21 has an initial stellar mass of about 950~M$_{\odot}$. We find that ASCC~19 and ASCC~21 have lost roughly 34\% and 28\% of their stellar mass, respectively, compared to their present-day stellar masses (796.85~M$_{\odot}$ and 686.09~M$_{\odot}$).

Since these two clusters have evolved for approximately 8.9~Myr, it is evident from the lower right sides of the two panels of Fig.~\ref{imf} that a portion of the massive stars has been depleted. Consequently, the loss of completeness in the current member stars of both clusters is likely more pronounced among the less-massive stars. However, as indicated by the low-mass ends of Fig.~\ref{imf}, the numbers of low-mass member stars lost are relatively small, suggesting that the completeness of the current member stars in our sample remains relatively high.

\end{appendix}

\end{document}